# Comparing weighting and imputation methods for enhancing statistical inference of health surveys given administrative claims data

Seho Park[*,1], Brianna L. Hardy[2], and A. James O'Malley[2, 3]

[1] Department of Industrial and Data Engineering, Hongik University, Seoul, South Korea
[2] The Dartmouth Institute for Health Policy and Clinical Practice, Geisel School of Medicine at Dartmouth, Lebanon, NH, USA
[3] Department of Biomedical Data Science, Geisel School of Medicine at Dartmouth, Lebanon, NH, USA

National surveys of the healthcare system in the United States were conducted to characterize the structure of healthcare system and investigate the impact of evidence-based innovations in healthcare systems on healthcare services. Administrative data is additionally available to researchers raising the question of whether inferences about healthcare organizations based on the survey data can be enhanced by incorporating information from auxiliary data. Administrative data can provide information for dealing with under-coverage-bias and non-response in surveys and for capturing more sub-populations. In this study, we focus on the use of administrative claims data to improve estimates about means of survey items for the finite population. Auxiliary information from the claims data is incorporated using multiple imputation to impute values of non-responding or non-surveyed organizations. We derive multiple versions of imputation strategy, and the logical development of methodology is compared to two incumbent approaches: a naïve analysis that ignores the sampling probabilities and a traditional survey analysis weighting by the inverses of the sampling probabilities. , and illustrate the methods using data from The National Survey of Healthcare Organizations and Systems and The Centers for Medicare & Medicaid Services Medicare claims data to make inferences about relationships of characteristics of healthcare organizations and healthcare services they provide.

## 1 Introduction

As large data sources have been widely available in many forms, statistical inferences based on survey data have new opportunities to be enhanced by combining survey data with information from other sources (Lohr and Raghunathan, 2017). Automated data sources such as electronic health records, claims and registries have been increasingly used as a tool for public health surveillance and collecting outcome information (Shortreed *et al.*, 2019; Kim and Shankar, 2020). This study is motivated by The National Survey of Healthcare Organizations and Systems (NSHOS), a large survey conducted on healthcare delivery organizations to investigate the impact of evidence-based innovations in healthcare systems. The NSHOS is a nationally representative survey of healthcare delivery systems in the United States fielded from June 2017 to August 2018. The goal of the survey was to characterize the structure of the healthcare system and to study the use of evidence-based innovations in healthcare systems and their impact on healthcare quality, delivery, and costs (Fisher *et al*., 2020). The U.S. health care system is well known for its complexity as the health organizations may be nested within larger health systems, and it leads to a multiple-tiered ownership structure as depicted in Figure 1.

---

[*]Corresponding author: e-mail: seho.park@hongik.ac.kr, Phone: +82-02-320-1606

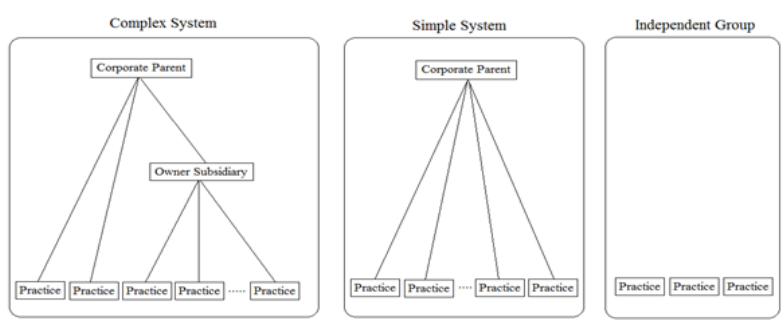

**Figure 1**  Data structure of healthcare organizations in the NSHOS survey.

An important feature of this survey is that a sampling frame of all healthcare organizations of each type (corporate parent, owner subsidiary, physician practice) used by researchers to design the survey is known at the time of analyzing the survey. Therefore, all variables used to form the weights for the survey for all organizations, including those unsampled, are available to us along with several additional variables that may be correlated with variables measured in the survey. The traditional method of using survey weights in estimation is to use the inverse of their selection probabilities as survey weights to inflate the importance of observations from the responding units so that they represent the population comprising all of non-respondents and non-surveyed units as well as the respondents (Pfeffermann, 1993, 1996; Pfeffermann *et al*., 1998; Skinner, 1994). As we have access to all the information in the sampling frame and additional auxiliary information beyond that, we consider the use of multiple imputation to form multiple complete data versions of a hypothetical survey of the entire sampling frame that had 100% response.

In addition to the NSHOS data, The Centers for Medicare & Medicaid Services (CMS) Medicare claims data is also available for all beneficiaries who receive medical care from the systems and practices in the sampling frame, including information on diagnoses, treatments, billed and paid amounts, insurance information, etc. Aggregated to the system or practice level, these variables may be strongly related to the study variables of interest. If so, they are useful for predicting what the survey responses would have been for units that were not surveyed or didn't respond. The Medicare claims data and the responses in the survey sample can be thought of as forming a predictive model that allows missing survey responses to be imputed. Imputations are performed multiple times to allow the uncertainty in the imputations to be accounted.

In general, imputation-based approaches are beneficial in terms of creating estimates based on combining information from multiple data sources with many strategies having been used (Durrant, 2009; Andridge and Little, 2010; Carpenter and Kenward, 2012). The primary advantage of imputation is that it provides complete datasets without missingness (Rubin, 1987, 1988, 1996), and may retain additional information for analysis than naïve case-deletion methods. If multiply imputed data sets are published, data users who only have limited access to publicly shared data can estimate models and perform analyses without requiring specific knowledge to address the missing data problems. In other words, each completed data set (data sets with the missing values imputed) can be analyzed using statistical methods for the analysis of unweighted data. However, the performance of multiple imputation will depend on the extent to which the variables are associated. An imputation model for multiple imputation is constructed based on multivariate relationships of variables. The stronger the associations, the more precise the missing data imputation and the less that uncertainty in the missing values will inflate standard errors and widen confidence intervals. The availability of auxiliary data for all organizations, sampled and unsampled, has the potential to enhance the performance of multiple imputation through more precise prediction of missing values.

The focus of this paper is the comparison of the analysis of the NSHOS survey data using traditional methods (naïve unweighted and sampling probability weighted) to a multiple imputation-based approach applied to the linked NSHOS survey data and the CMS Medicare claims data that utilizes auxiliary data and avoids using survey weights. To motivate and illustrate the methodology, we focus on the construction of simple inferences (e.g., population totals and averages) for the population of physician practices represented by the NSHOS survey. We will compare existing approaches to our novel multiple imputation-based approach that utilizes auxiliary information from CMS Medicare claims data. Compared to weighted estimation, which is commonly used for handling data that are subject to missingness (Pfeffermann, 1993; Pfeffermann *et al*., 1998; Wiśniowski *et al*., 2020), multiple imputation also provides an appropriate way of creating complete datasets for subsequent statistical analyses for multiple purposes and by multiple users. We demonstrate that the multiple imputation approach can improve accuracy of estimation in comparison with other traditional estimation methods used for survey data. We illustrate these methods analyzing the NSHOS survey data collected in 2017 and auxiliary CMS Medicare claims data from 2015.

In Section 2, the multiple imputation procedure for multilevel data used in the study is developed. In our statistical analyses we wish to respect the statistical dependence structure of our data to ensure that all analyses of the resulting data sets, including unadjusted means and totals, are correctly calibrated in terms of statistical precision. In Section 4, we describe the NSHOS survey data in greater detail, and in Section 3, descriptive inference for the parameters of the finite population represented by the sampling frame is developed. In Chapter 5, we evaluate and compare statistical inferences for the NSHOS study using naïve (complete-case) analysis, sampling weights, and multiple imputation. In Chapter 6, we summarize the methods and findings, and review the advantages and disadvantages of the methods.

## 2 Multiple imputation applied to a sampling frame given survey data

Given responses of the NSHOS survey sample, sampling probability weighted estimation can provide point and interval estimators that possess desired properties such as unbiasedness and statistical efficiency (Pfeffermann, 1993, 1996; Skinner, 1994; Fuller, 2009; Rao *et al*., 2010). The crucial feature of this study is that the sampling frame used for the NSHOS survey is available for data analyses. Therefore, some variables such as number of hospitals or owner subsidiaries in the system, or primary care physicians, etc., known prior to the survey, were measured for all units in the sampling frame. Auxiliary information for the units not selected for the sample or non-respondents to the survey is not included in weighted estimation, which focuses entirely on the surveyed units. In addition to the auxiliary information in the sampling frame, other external sources of information may also be available to expand our knowledge of the unsampled units and the relationships between variables of primary interest (the survey variables) and auxiliary variables. In our case, the CMS Medicare claims data are available as additional information to aid estimation of practice and system-level means. As an alternative to using survey weights, we consider using an imputation method that can provide researchers complete and comprehensive datasets for an entire sampling frame without missingness and without having surveyed most units.

A sampling unit of the NSHOS survey is a physician practice with 3 or more primary care physicians defined as family medicine, geriatrics, internal medicine, or preventive medicine specialties. The NSHOS implemented stratified cluster sampling to collect samples of physician practices that have different ownership and composition structures; they may be either system-owned or independent. Given the sampled physician practices from the NSHOS survey, we link the physician practices to the CMS claims data for beneficiaries who received medical care from the practices in order to obtain additional information from auxiliary data.

Claims data on Medicare beneficiaries were matched to the physician practice sampling frame data by exploiting the fact that individuals' claims have Tax Identification Numbers (TIN). The TIN that was used most often per beneficiary in the claims data was pulled into the attribution algorithm (beneficiaries to assigned TIN). These same TINs may be linked to OneKey data using the MD-PASS national provider identifier to TIN crosswalk to identify the practice based on where the plurality of their care is performed. Primary care providers were favored over specialty providers when taking into consideration the plurality of care (Centers for Medicare & Medicaid Services, 2015, 2018). The integrated data sources combined to form a rich data set consisting of study variables, other variables from the NSHOS survey sample, auxiliary variables from the sampling frame, and additional variables from the CMS Medicare claims data.

Multiple imputation is one of the most popular approaches to protecting against loss of information and estimation bias due to missing data (Rubin, 1984, 1996). It is reliant on the data being missing at random (MAR), implying that while the probability of missing data on a variable may be related to measured variables, it has no dependence on any unmeasured values of any variables. For multilevel data, missingness can be observed at multiple levels. Complications for performing multiple imputation arise with multi-level data because it is important that the imputed data retain the within-group correlation structure of the original data, otherwise biases and statistical inefficiency in estimation (Goldstein, 1995; 1998, Grund *et al*., 2016, 2018a, 2018b; Hox and Roberts, 2010) may result. In our study, the NSHOS data have missing entries at the practice-level and the system-level (both owner subsidiary and corporate parent). We used a missing-data imputation model for multivariate data that preserves the multi-level structure of the data by constraining the imputations to have the same within-group correlations at levels 2 and 3 as in the original data (see Yucel (2008) for details). The method is implemented in `the R package PAN` (Yucel, 2008). Missing data imputation in `R` is supplemented with our use of a `SAS` procedure for performing weighted mean estimation with the weights treated as probability weights.

## 3    Description of the NSHOS survey data

### 3.1    Sampling design and response rate of the NSHOS survey

Three different clustering sampling designs, ranging from three-level to single level, are implemented to extract the sampling units for the NSHOS survey (Figure 2).

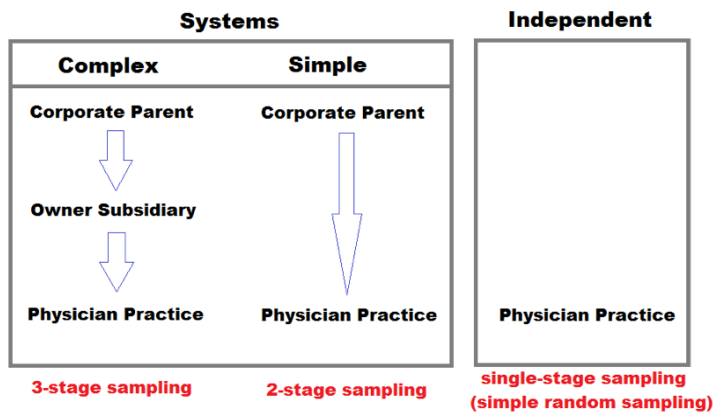

**Figure 2**    Sampling designs used for the NSHOS survey.

The NSHOS survey data were collected between June 2017 and August 2018. Systems owned by the federal government and systems with a reference to specialty in their business name were excluded. For healthcare systems in the NSHOS survey, 341 out of 570 systems (59.8%) responded. After 16 ineligible responses were excluded from the responses, 325 responses were analyzed. For owner subsidiary in the NSHOS survey, 107 out of 222 owner subsidiaries responded (48.2%) and 8 ineligible responses were excluded. In total 99 responses were analyzed. System-associated physician practices were selected and target respondents were individuals able to speak for the practice (e.g., medical director, practice manager). For physician practices in the NSHOS survey, 2,333 of 4,976 physician practices (46.9%) responded, and after removing ineligible responses, 2,190 responses were analyzed.

## 4    Descriptive statistics for the NSHOS survey

### 4.1    Mean estimation incorporating sampling weights

A traditional cluster or stratified sampling design for a survey is likely to be a complex survey design for which methods that account for the dependencies in the sampling weights (design-based inference) or the survey items (model-based inference) need to be used. Even when a model-based approach is being used, ignorance of sampling weights at each level of data leads to substantial biases in estimation of finite population parameters (Pfeffermann 1994; Pfeffermann et al., 1998; Skinner 1989; Fuller, 2009; Rao et al., 2010). We follow the general strategy of accounting for sampling weights when seeking to estimate a hierarchical model by including sampling weights equal to the reciprocal of the selection probabilities at that level (Pfeffermann et al., 1998; Pfeffermann et al., 1998, 2006; Rabe-Hesketh and Skronda, 2006; Asparouhov and Muthen, 2006, 2008).

When estimating a simple population feature such as mean, total, or proportion of the variable of interest denoted θ, the weighted estimator when more than one-level of hierarchy exists in the data is written as:

$$\hat{\theta} = \frac{\sum_i \sum_j \sum_k w_{ijk} y_{ijk}}{\sum_i \sum_j \sum_k w_{ijk}},$$

where the sampling weight, $w_{ijk}$, is a product of survey weights at each level. This is the classic Horvitz-Thompson (HT) estimator (Horvitz and Thompson, 1952) of a population mean that sets the sampling weights equal to a reciprocal of a corresponding selection probability of the unit. In the NSHOS data, we have three types of health organizations comprising a three-level hierarchy with sampling weights reflecting the use of sampling at each stage. Let $\pi_{ijk}$ be a selection probability of the $k$-th physician practices within the $j$-th owner subsidiary that is nested within the $i$-th corporate parent. Then we can write the selection probability as

$$\pi_{ijk} = \pi_i \times \pi_{j|i} \times \pi_{k|ij} = P(i \in C) \times P(j \in O | i \in C) \times P(k \in P | i \in C, j \in O)$$

where $C, O, P$ represent a group of units at the corporate parent, owner subsidiary, and physician practice levels, respectively. Thus, the survey weight $w_{ijk}$ is a reciprocal of the selection probability $\pi_{ijk}$, and is also called a design weight. In the NSHOS survey, the selection probabilities at each of the three levels are all available and they are generated by a nonlinear function of measures in the sampling frame. More details of the procedure are in O'Malley and Park (2020).

A variance estimator of the HT estimator is constructed using design-based inference and so depends on the sampling design used in the data collection. In the NSHOS survey, three-stage cluster sampling was implemented and so the variance of the HT estimator in this study can be expressed as:

$$\widehat{\text{Var}}(\hat{\theta}) = \text{Var}\big[\mathbf{E}\{\mathbf{E}(\hat{\theta}|s_1, s_2)|s_1\}\big] + \mathbf{E}\big[\text{Var}\{\mathbf{E}(\hat{\theta}|s_1, s_2)|s_1\}\big] + \mathbf{E}\big[\mathbf{E}\{\text{Var}(\hat{\theta}|s_1, s_2)|s_1\}\big], \qquad (1)$$

where $s_1$ and $s_2$ represent a sample of primary sampling units and secondary sampling units, respectively. It is well established that the design-based variance estimator is unbiased for the variance of the HT estimator: $\mathbf{Var}(\hat{\theta})$.

### 4.2    Mean estimation after multiple imputation

Multiple imputation applied to multilevel data should take account of dependencies in the data and allows relationships between variables to be estimated at different levels of analysis or effects that may vary across higher-level observational units. Under the assumption of MAR, auxiliary variables from the CMS Medicare claims are included in the imputation model along with any available variables that are suspected of being related to either the propensity of missing data or the missing values themselves (Collins, Schafer and Kam, 2001). Once multiple imputed datasets are constructed after the multiple imputation, we obtain a mean estimator after multiple imputation:

$$\bar{y}_{MI} = \frac{1}{D} \sum_{d=1}^{D} \bar{y}_d,$$

where $d$ is an index for imputation number: $\{d = 1, \cdots, D\}$. In each imputation $d$, we obtain a mean estimator from the $d$-th imputed dataset, denoted $\bar{y}_d$, and the average of the $D$ imputed estimates, $\bar{y}_d$, is a multiple imputation mean estimator. By Rubin's rules (Rubin, 1987), the variance estimator of the multiple imputation mean estimator is given by

$$\mathbf{Var}_{total} = \mathbf{Var}_w + \left(1 + \frac{1}{D}\right)\mathbf{Var}_B, \qquad (2)$$

where

$$\mathbf{Var}_w = \frac{1}{D}\sum_{i=1}^{D}\mathbf{Var}_d$$

and $\mathbf{Var}_d$ is a variance estimator for population mean evaluated on the $d$-th imputed dataset, and

$$\mathbf{Var}_B = \frac{1}{D-1}\sum_{i=1}^{D}(\bar{y}_d - \bar{y})^2.$$

## 5 Data Application

The R package PAN uses the joint modeling approach for performing multiple imputation such that all incomplete variables are simultaneously imputed using a single statistical model (Schafer & Yucel, 2002). This preserves the features and characteristics of the multilevel data structure and it is crucial to obtain reliable estimates from multilevel data analysis (Grund *et al*., 2016, 2018b). But a drawback of the joint modeling approach is that it is usually confined to random intercept analyses and not applicable when random slope variables contain missingness (Enders *et al*., 2016; Speidel *et al*., 2018).

As we incorporate the full sampling frame into the analysis, missing values are observed in predictor variables as well as outcome variables. The missingness may occur due to the non-sampled units in the sampling frame in addition to the non-responses in the survey. In order to handle missing predictor variables, several approaches are suggested: Performing multiple imputation separately for each group using single-level multiple imputation (Graham, 2012) or treating incomplete predictor variables as outcome variables in the imputation model (Schafer, 2001). We follow Schafer's approach to generate imputations for incomplete predictor variables prior to implementing multiple imputation for outcome variables, which may generate a bias for the slope variance, but theory and past results suggest this is likely to be small and acceptable (Wu, 2010; Goldstein *et al*., 2014; Grund *et al*., 2016). Thus, we first generate imputed values for creating complete predictor variables data and then implement multiple imputation for outcome variables using the complete predictor variables.

### 5.1 Imputed Covariates

In order to build an imputation model for generating imputed values for incomplete predictor variables, we needed variables that were known across the sampling frame. We chose six variables from the sampling frame, which conveniently were not subject to missingness, to be used for generating synthetic imputed values for those variables that were missing either due to not being sampled or were sampled but did not respond. The six variables are given in Appendix A1.

After implementing multiple imputation with $D=10$, the imputed values for the missing values of the predictor variables yield 10 data sets with complete data. By using the complete predictor variables where the missing values are filled in by an average of 10 imputed values, we build an imputation model for the outcome variables and generate imputed values for missing outcome variables to create 10 complete datasets.

## 5.2 Point Estimators

The NSHOS survey seeks to estimate characteristics of healthcare systems and impacts of innovations for the healthcare services in the United States. In order to understand the features, we are interested in estimating the population mean or proportion of characteristics of healthcare services at physician practices. The mean after multiple imputation is compared with two other types of point estimators: a naïve mean estimator and a weighted mean estimator (HT estimator).

1) Naïve mean estimator:

$$\bar{y}_{naive} = \frac{1}{n}\sum_i \delta_i y_i,$$

where $= 1$ when the $i$-th unit of $y$ is observed or 0 otherwise, and $n$ is the number of observations in the dataset ($\sum_i \delta_i = n$). The naïve mean performs case-deletion with non-responding and non-selected units excluded from the analysis. The variance estimator of the naïve mean is given by $\mathbf{Var}(\bar{y}_{naive}) = \frac{1}{n}\mathbf{Var}(y) = \frac{1}{n}\sigma_y^2$, which equals the square of the standard error of the mean.

2) Weighted mean estimator:

$$\bar{y}_w = \frac{\sum_i \delta_i w_i y_i}{\sum_i \delta_i w_i},$$

where $w_i$ is a sampling weight corresponding to the observed value $y_i$. The weighted mean estimator considers completely measured observations from responses in the survey sample and estimates a finite population mean by inflating the observations' representativeness to the population size using sampling weights. A design-based variance estimator of the weighted mean estimator is computed using the formula given in (1).

3) Mean estimator after multiple imputation:

$$\bar{y}_{MI} = \frac{1}{D}\sum_{d=1}^{D} \bar{y}_d,$$

where $\bar{y}_d$ is a mean estimator from the $d$-th imputed dataset, and $\bar{y}_{MI}$ is an average of the mean estimators from D imputed datasets. A variance estimator of the multiple imputation mean estimator is obtained by Rubin's rules given in (2).

For multiple imputation estimation, two types of multiple imputation scenarios are considered in order to demonstrate how the effectiveness of the procedure is enhanced by having more information with which to impute the missing values. In our case the auxiliary information comes from the aggregated CMS Medicare claims data. The two scenarios are:

1) Using covariates from the sampling frame (MI1): The six variables in Table 4 in Appendix A1 are used as covariates in the imputation model. All the variables are completely observed in the sampling frame and thus they are available for direct multiple imputation for outcome variables.

2) Using the CMS Medicare claims data-based covariates (MI2): Additional variables from the CMS Medicare claims data are listed in Table 5 in Appendix A1. As the CMS Medicare claims data are observed at the beneficiary level, where a beneficiary is considered as a component of a practice, practice-level means (or proportions in the case of binary variables) of the variables are formed.

## 5.3 Results

We report a result of physician practice-level variables in this section and other level's results are given in Table 6 and Table 7 in Appendix A2.

**Table 1** Estimated mean and standard errors of physician practice-level variables: naïve estimator (Naive); Weighted estimator (Weighted); Multiple imputation using covariates from the sampling frame (MI1); Multiple imputation using covariates from the CMS Medicare claims data (MI2)

|  | Naïve | | Weighted | | MI1 | | MI2 | |
|---|---|---|---|---|---|---|---|---|
|  | Mean | S.E. | Mean | S.E. | Mean | S.E. | Mean | S.E. |
| FQHC† | 0.147 | 0.010 | 0.148 | 0.013 | 0.192 | 0.010 | 0.196 | 0.007 |
| % APM‡ participation | 0.808 | 0.011 | 0.808 | 0.012 | 0.797 | 0.010 | 0.797 | 0.0089 |
| % Medicare ACO§ participation | 0.504 | 0.014 | 0.502 | 0.019 | 0.506 | 0.013 | 0.506 | 0.011 |
| % Medicaid ACO participation | 0.367 | 0.014 | 0.363 | 0.015 | 0.387 | 0.010 | 0.388 | 0.009 |
| % Common ACO participation | 0.454 | 0.014 | 0.447 | 0.018 | 0.465 | 0.014 | 0.461 | 0.011 |
| % Any ACO participation | 0.648 | 0.014 | 0.646 | 0.017 | 0.618 | 0.010 | 0.618 | 0.010 |
| # Emergency department visits per 10K | 0.683 | 0.008 | 0.686 | 0.011 | 0.683 | 0.010 | 0.683 | 0.009 |
| % Partial dual | 0.047 | 0.002 | 0.047 | 0.003 | 0.052 | 0.002 | 0.052 | 0.001 |
| % Full dual | 0.181 | 0.005 | 0.184 | 0.009 | 0.189 | 0.004 | 0.185 | 0.003 |
| Mean # HCC§§'s/Patient | 2.419 | 0.020 | 2.428 | 0.027 | 2.409 | 0.018 | 2.398 | 0.019 |
| Mean # Admissions per 100 | 0.760 | 0.031 | 0.764 | 0.035 | 0.847 | 0.025 | 0.843 | 0.023 |
| Mean % depression diagnosis | 0.202 | 0.002 | 0.202 | 0.003 | 0.202 | 0.002 | 0.202 | 0.002 |
| Mean % SMI* diagnosis | 0.045 | 0.002 | 0.045 | 0.002 | 0.050 | 0.002 | 0.050 | 0.001 |
| Mean % mental health need | 0.279 | 0.003 | 0.280 | 0.004 | 0.279 | 0.003 | 0.283 | 0.002 |

†Federally Qualified Health Center; ‡Alternative Payment Models; §Accountable Care Organizations; §§Hepatocellular carcinoma; *Serious Mental Illness

Table 1 contains four estimated means and associated standard errors of the characteristic variables of healthcare services measured at physician practice-level in the NSHOS survey. It is well established that a weighted mean estimator is unbiased for the finite population mean as long as design weights take into account the complex survey design used for the sample collection and the MAR assumption is satisfied in the dataset. We compare the estimators using the weighted estimator as a reference for population mean estimation and compare other mean estimators against it to detect whether it is biased. Considering weighted mean estimates as the reference, naïve mean estimates tend to be smaller or equal to the weighted estimates, whereas mean estimates after multiple imputation are greater than or equal to the weighted estimates. While similar to mean estimates obtained from MI2, estimates obtained from MI1 are sometimes less than mean estimates of MI2. For example, an outcome variable of 'mean number of admissions per 100' has mean estimates of 0.760, which is less than the weighted estimate of 0.764 and two mean estimates after multiple imputation – MI1 and MI2 - are 0.847 and 0.843, respectively, which are slightly larger than the weighted estimate. The mean estimate of MI2 is closer to the weighted mean than the mean estimate of MI1, implying that the imputation model including predictor variables from the CMS Medicare claims data provides more predictability with less estimation bias.

Notably, standard errors of the mean estimators vary markedly across the four different types of estimation. In most cases, MI2 has the smallest standard errors whereas weighted estimation has the largest standard errors. For the 'mean number of admissions per 100' variable, the standard error of weighted estimation, 0.035, is larger than the standard error of naïve estimation, 0.031, and is smaller than standard errors of the two multiple imputation scenarios (MI1 and MI2), 0.025 and 0.023, respectively. The standard error of the weighted estimator is naturally larger than the standard error of a naïve estimator because sampling weights inflating representativeness of each unit are incorporated in the variance estimation (the variance of the estimator typically increases with the variance of the weights).

In contrast, there is a decrease in the standard error of the mean estimator when we use multiple imputation. Although we implemented a two-step imputation procedure (for predictor variables and then outcome variables), which induces the additional imputation variation accounted for by the inflation factor defined by Rubin's rules given in equation (2), the standard errors resulting from the two multiple imputation scenarios MI1 and MI2 were smaller than the standard error of the traditional survey weighted estimation. This finding supports the use of auxiliary information, specifically the incorporation of variables related to the outcome variable, in the imputation model.

The above findings can be explained by 1) an increased sample size due to the imputed values, which increased from 1,208 to 2,702 following multiple imputation, and by 2) the additional information obtained from the auxiliary data that increases the precision of the multiply imputed values. Furthermore, MI2 has a smaller standard error compared to MI1, implying that including the covariates derived from the CMS Medicare claims data better inform the population mean. Although there is a penalty due to the uncertainty in the imputed values, which increases the computed variance via Rubin's rules, this is offset by decreased due to the auxiliary information from the CMS Medicare claims data yielding imputed values whose standard error of prediction is relatively small.

The following two tables, Table 2 and Table 3, examine the efficiency of using multiple imputation with the two different sets of variables underlying MI1 and MI2. In order to check whether the gain in efficiency is due solely to the increased sample size in the completed datasets post imputation, we will evaluate and compare both sides of the following equation:

$$\frac{S.E. \text{ of each imputed dataset}}{S.E. \text{ of the sample}} = \sqrt{\frac{n_{original}}{n_{imputed}}}. \qquad (3)$$

The left hand of the equation is the ratio of two standard errors: one computed from a complete dataset created after each imputation procedure (the numerator) and one computed from the original incomplete dataset (the denominator). The right-hand side of the equation is the ratio of two sample sizes: one observed from original sample ($n_{original}$) as a numerator and one observed from the dataset after imputation ($n_{imputed}$) as a denominator. If the two sides are equal so (3) holds, it implies that the decrease in the standard error of the multiple imputation-based estimator is primarily due to the increase of the sample size, implying that the correlations among the variables are sufficiently great that they entirely offset the penalty factor (in Rubin's rule).

The outcome variable used as an example is 'percentage of APM participation at physician practices.' The sample size of the NSHOS survey is $n_{original}$=1,208 and the sample size after imputation is $n_{imputed}$=2,702, so the square root of the ratio of two sample sizes is 0.669. We compare this value to the ratio of two standard errors in Equation (3). The standard error of the variable in the NSHOS sample is 0.011. The same for the multiple imputation procedure applied to both sets of variables used for multiple imputation is now computed.

We report 10 estimates from 10 imputed datasets and its ratio that is defined by standard error of the estimate from imputed dataset divided by standard error of the estimate from original sample.

A. Multiple imputation using covariates from the sampling frame (MI1)

**Table 2**   MI1: Estimates from the 10 Imputed datasets of the outcome variable

|       | $Y_1$ | $Y_2$ | $Y_3$ | $Y_4$ | $Y_5$ | $Y_6$ | $Y_7$ | $Y_8$ | $Y_9$ | $Y_{10}$ |
|-------|-------|-------|-------|-------|-------|-------|-------|-------|-------|----------|
| Mean  | 0.802 | 0.802 | 0.807 | 0.792 | 0.807 | 0.806 | 0.804 | 0.797 | 0.805 | 0.808    |
| S.E.  | 0.006 | 0.006 | 0.006 | 0.006 | 0.006 | 0.006 | 0.006 | 0.006 | 0.006 | 0.006    |
| Ratio | 0.583 | 0.581 | 0.569 | 0.586 | 0.572 | 0.570 | 0.577 | 0.589 | 0.574 | 0.573    |

Since we used 10 as the number of imputations, 10 evaluations of the ratio are reported in Table 2. The two sources of the variance estimate of multiple imputation, which are the square root of the variance within an imputed dataset ($\sqrt{Var_w}$) and the square root of the variance between imputed datasets ($\sqrt{Var_b}$), are 0.006 and 0.007, respectively, when the standard error of the variable after MI1 is 0.010.

B. Multiple imputation using covariates from the CMS Medicare claims data (MI2)

**Table 3**   MI2: Estimates from the 10 Imputed datasets of the outcome variable

|       | $Y_1$ | $Y_2$ | $Y_3$ | $Y_4$ | $Y_5$ | $Y_6$ | $Y_7$ | $Y_8$ | $Y_9$ | $Y_{10}$ |
|-------|-------|-------|-------|-------|-------|-------|-------|-------|-------|----------|
| Mean  | 0.805 | 0.809 | 0.799 | 0.803 | 0.803 | 0.803 | 0.791 | 0.799 | 0.790 | 0.799    |
| S.E.  | 0.006 | 0.006 | 0.006 | 0.006 | 0.006 | 0.006 | 0.006 | 0.006 | 0.006 | 0.006    |
| Ratio | 0.575 | 0.568 | 0.587 | 0.576 | 0.578 | 0.585 | 0.559 | 0.600 | 0.599 | 0.588    |

In case of using MI2, $\sqrt{Var_w}$ and $\sqrt{Var_b}$ are 0.006 and 0.006, respectively, when the standard error of the variable after MI2 is 0.009.

From Table 2, we found that all 10 values of the left hand of Equation (3), the ratio of the standard error of each of the 10 imputed datasets to the standard error of the original incomplete dataset, is less than the right hand of the Equation (3), of 0.669. The same results can be found in Table 3. The results in Table 2 and Table 3 imply that the correlations among the variables are such that the decrease in the standard errors after imputation (MI1 and MI2) offset the multiple imputation penalty. In addition to the increase in sample size, the comparisons suggest that other advantageous features of multiple imputation are related to the improvement in variance estimation. Because the values of the left hand of the Equation (3) reported in Table 2 are much less than the values reported in Table 3, this suggests that the use of CMS Medicare claims data further enhances the predictive power among the variables used in the multiple imputation. The source of such precision is the correlation between the outcome variable and the CMS Medicare claims-derived means. Inspection of the correlations between the variables in the imputation model and outcome 'percentage of diagnosis of depression' are reported in the Appendix. Table 8 and Table 9 report a summary of magnitude of correlation between outcome variable and the covariates in the sampling frame and those obtained from the CMS Medicare claims data, respectively. It is clearly noticeable that covariates from CMS Medicare claims have correlations of greater magnitude on average with the outcome variable than those from the sampling frame: percentage of diagnosis of depression is highly associated with characteristics of patients such as mean of age, percentage of female, median household income, practice size, and percentage of Hispanic patients receiving care at the practice. It explains the improvement of estimation using MI2, which uses the CMS Medicare claims aggregate variables in the imputation model. This confirms that the higher the correlation between the CMS Medicare claims data variables, the more precise the imputations and the greater the reduction in the standard error, and it proves an advantage of these methods over competing methods in terms of enhanced performance in accuracy and efficiency of estimation.

## 6   Discussion

We compared three different methods for mean and variance estimation of parameters representing population means of items in the NSHOS survey sample data and demonstrated that the use of auxiliary data in the form of the CMS Medicare claims data yielded more precise estimators at all levels of data. By incorporating sampling weights that account for the complex and unique hierarchies in the NSHOS data and because they are well known for providing unbiased estimates under the MAR assumption, the weighted mean estimates are considered a reference for comparison with other methods. We demonstrated using the NSHOS survey data that combining auxiliary information from additional resources via multiple imputation appears on target (i.e., point estimates are close to the weighted estimator) while achieving much smaller variance estimates. We also demonstrated via the involvement of CMS Medicare claims data that the imputation procedure can be enhanced by using a set of variables for imputing missing values of the study variables that as a group are substantially correlated with the study variables. Therefore, our new procedure has the potential to enhance the analysis of survey data by yielding much more precise inferences than possible under traditional procedures based on weighting with the precision gain appearing proportional to the inverse of the sampling probabilities.

Most data users of surveys such as the NSHOS do not have access to the sampling frame for the study. They are only given access to use public-use data sets and survey weights shared by data developers or owners. Often the end users are distinct entities from the survey developers, so the full sampling frame is available to only a few survey data users. This may be a limitation of applying our multiple imputation-based approaches. However, with the ever increasing use of synthetic data, it is possible that multiply imputed data may become the norm for public-use data sets than the exception.

As an extension of our study, multi-level models can be implemented without needing to be concerned about involving sampling weights at each level and adjusting sampling weights for non-responses to avoid biases in estimation. Under the MAR assumption in the survey data and given the availability of the full sampling frame along with external sources of information, we can estimate much more complicated multi-level models on complete data sets. We do not need to repeat the imputation procedure as the same imputed data sets may be used as those that were used for determination of descriptive summaries. Rather, we simply substitute the results from the fitted multilevel model for the descriptive population mean estimators used for illustration in this paper. The resulting estimates can be more accurate than those obtained from analyses involving multi-level models estimated using survey weights, for which uncertainty remains as to exactly how to incorporate the survey weights nor of what values they should equal. In this sense, the multiple imputation approach developed above avoids having to come up with an ad hoc procedure or approximations such as those currently being used for weighted estimation of multi-level models.

**Acknowledgements** An acknowledgement may be placed at the end of the article. This work was supported by the Agency for Healthcare Research and Quality's (AHRQ's) Comparative Health System Performance Initiative under Grant # 1U19HS024075, which studies how health care delivery systems promote evidence-based practices and patient-centered outcomes research in delivering care. The findings and conclusions in this article are those of the authors and do not necessarily reflect the views of AHRQ. The data that support the findings of this study are available on request from the corresponding author. The data are not publicly available due to privacy or ethical restrictions. The statements, findings, conclusions, views, and opinions contained and expressed in this article are based in part on data obtained under license from IQVIA information services: OneKey subscription information services 2010-2018, IQVIA incorporated all rights reserved. The statements, findings, conclusions, views, and opinions contained and expressed herein are not necessarily those of IQVIA Incorporated or any of its affiliated or subsidiary entities. AMA is the source for the raw physician data; statistics, tables or tabulations were prepared by the authors using AMA Masterfile data.

**Conflict of Interest**
*The authors have declared no conflict of interest*

# Appendix

## A.1 Covariate Variables from the Sampling frame and the CMS Medicare claims data

**Table 4** Six variables measured completely from the sampling frame.

| Variable | Description |
|---|---|
| np | Number of physicians in organization |
| npcp | Number of primary care physicians in organization |
| nach | Total number of hospitals in system |
| nmg | Total number of medical groups in system |
| nos | Number of owner subsidiaries in system |
| pertot | Proportion of hospitals and medical groups under local ownership unit (i.e., not counting those under other ownership structures within the system) |

**Table 5**  Variables from the CMS Medicare claims data.

| Variable | Description |
|---|---|
| Practice size | Number of patients per practice |
| Region | Region of practice (Midwest/Northeast/South/West) |
| Ruca | Proportion of rural beneficiaries |
| Mean of age | Integer age at beginning of year |
| Percentage of female | Proportion of female beneficiaries |
| Mean of median household income | Mean of median household income using 2010 census TMHI[†] and 2016 zip codes |
| System size | System size by number of hospitals |
| Percentage of white | Proportion of white beneficiaries |
| Percentage of black | Proportion of black beneficiaries |
| Percentage of Hispanic | Proportion of Hispanic beneficiaries |
| Percentage of other | Proportion of other beneficiaries |

[†]TMHI: Median household income in the past 12 months

### A.2 Results for owner subsidiary and corporate parent level variables

1) Owner subsidiary-level variables

**Table 6**  Estimated mean and standard errors of owner subsidiary-level variables: Naive, Weighted, Multiple Imputation 1 (MI1) using covariates from the sampling frame, Multiple Imputation 2 (MI2) using covariates from the CMS Medicare claims data (o.s: owner subsidiary)

| | Naïve | | Weighted | | MI1 | | MI2 | |
|---|---|---|---|---|---|---|---|---|
| | Mean | S.E. | Mean | S.E. | Mean | S.E. | Mean | S.E. |
| Number of acute care hospitals under o.s | 2.843 | 0.370 | 3.300 | 0.949 | 4.035 | 0.488 | 3.887 | 0.448 |
| Number of states o.s operates in | 1.560 | 0.106 | 1.666 | 0.289 | 1.722 | 0.149 | 1.742 | 0.106 |
| Flag if o.s is part of an ACO | 0.076 | 0.008 | 0.082 | 0.019 | 0.116 | 0.006 | 0.117 | 0.005 |
| Number of critical access hospitals under o.s | 0.678 | 0.171 | 0.765 | 0.426 | 1.498 | 0.191 | 1.588 | 0.144 |
| Number of psychiatric hospitals under o.s | 0.103 | 0.016 | 0.106 | 0.028 | 0.219 | 0.026 | 0.200 | 0.021 |
| Number of rehabilitation hospitals under o.s | 0.095 | 0.018 | 0.089 | 0.030 | 0.203 | 0.041 | 0.165 | 0.019 |
| Number of practices under o.s | 38.183 | 1.621 | 37.986 | 3.730 | 40.370 | 3.920 | 37.506 | 2.248 |
| Number of practices with more than 3 PCP†s in the o.s | 12.185 | 0.513 | 11.860 | 1.028 | 11.566 | 0.578 | 11.611 | 0.580 |
| Mean number of physicians per practice under o.s | 7.163 | 0.419 | 7.167 | 0.904 | 7.138 | 0.903 | 8.451 | 0.603 |
| Number of practices with 1 physician under o.s | 7.268 | 0.419 | 7.322 | 1.029 | 8.635 | 1.016 | 7.824 | 0.614 |
| Number of practices with 2-9 physicians under o.s | 24.480 | 1.064 | 24.185 | 2.403 | 25.221 | 2.618 | 24.395 | 1.983 |
| Number of practices with 10-20 physicians under o.s | 3.548 | 0.212 | 3.598 | 0.459 | 3.836 | 0.427 | 3.565 | 0.211 |
| Number of practices with 21+ physicians under o.s | 1.963 | 0.207 | 2.027 | 0.610 | 2.629 | 0.359 | 2.380 | 0.193 |

†Primary Care Physicians

2) Corporate parent-level variables

**Table 7** Estimated mean and standard errors of corporate parent-level variables: Naïve, Weighted, Multiple Imputation 1(MI1) using covariates from the sampling frame, Multiple Imputation 2 (MI2) using covariates from the CMS Medicare claims data (c.p: corporate parent)

| | Naïve | | Weighted | | MI1 | | MI2 | |
|---|---|---|---|---|---|---|---|---|
| | Mean | S.E. | Mean | S.E. | Mean | S.E. | Mean | S.E. |
| Number of acute care hospitals under c.p | 16.128 | 0.941 | 17.178 | 3.621 | 17.787 | 0.859 | 17.012 | 0.630 |
| Number of states c.p. operates in | 3.672 | 0.169 | 3.888 | 0.575 | 3.926 | 0.120 | 3.862 | 0.119 |
| Flag if c.p. is part of an ACO | 0.355 | 0.014 | 0.353 | 0.031 | 0.366 | 0.010 | 0.365 | 0.009 |
| Number of critical access hospitals under c.p | 3.180 | 0.220 | 3.388 | 0.807 | 3.606 | 0.165 | 3.583 | 0.150 |
| Number of psychiatric hospitals under c.p | 1.225 | 0.222 | 1.274 | 0.433 | 1.425 | 0.325 | 1.427 | 0.363 |
| Number of rehabilitation hospitals under c.p | 0.614 | 0.042 | 0.668 | 0.157 | 0.712 | 0.031 | 0.695 | 0.029 |
| Number of practices under c.p | 117.805 | 5.273 | 123.394 | 20.493 | 132.870 | 4.534 | 127.837 | 3.957 |
| Number of practices with 3+ PCPs in the c.p | 33.037 | 1.471 | 34.444 | 5.806 | 35.584 | 1.170 | 35.419 | 1.229 |
| Mean number of physicians per practice under c.p | 6.365 | 0.147 | 6.366 | 0.444 | 6.394 | 0.102 | 6.372 | 0.101 |
| Number of practices with 1 physician under c.p | 25.638 | 1.332 | 27.061 | 5.149 | 28.945 | 1.113 | 28.469 | 0.950 |
| Number of practices with 2-9 physicians under c.p | 72.314 | 3.276 | 75.592 | 12.699 | 79.794 | 2.924 | 78.316 | 2.591 |
| Number of practices with 10-20 physicians under c.p | 10.030 | 0.468 | 10.481 | 1.810 | 10.879 | 0.341 | 10.775 | 0.320 |
| Number of practices with 21+ physicians under c.p | 6.172 | 0.625 | 6.468 | 2.693 | 7.037 | 0.621 | 7.069 | 1.065 |

### A.3 Correlation between variables and outcome variable

1) Summary of magnitude of correlation between covariates from sampling frame and outcome variable ($Y$ = percentage of diagnosis of depression)

**Table 8** Summary of magnitude of correlation between covariates from sampling frame and an outcome variable.

| Variable | $Y$ | Minimum | Mean | Maximum | St. Dev |
|---|---|---|---|---|---|
| np | 0.003 | 0.000 | 0.008 | 0.022 | 0.007 |
| npcp | 0.052 | 0.027 | 0.060 | 0.090 | 0.018 |
| nach | 0.025 | 0.004 | 0.031 | 0.075 | 0.025 |
| nmg | -0.001 | 0.001 | 0.015 | 0.049 | 0.013 |
| nos | 0.011 | 0.003 | 0.021 | 0.073 | 0.023 |
| Pertot | 0.067 | 0.013 | 0.047 | 0.084 | 0.020 |

Description of the variables listed in the Table 8 is reported in the Table 4 in the Appendix A1.

2) Summary of magnitude of correlation between covariates from the CMS Medicare claims data and outcome variable ($Y$ = percentage of diagnosis of depression)

**Table 9** Summary of magnitude of correlation between covariates from CMS Medicare claims data and an outcome variable.

| Variable | $Y$ | Minimum | Mean | Maximum | St. Dev |
|---|---|---|---|---|---|
| Practice size | 0.074 | 0.018 | 0.082 | 0.165 | 0.042 |
| Region | -0.068 | 0.016 | 0.049 | 0.085 | 0.021 |
| Ruca | -0.006 | 0.002 | 0.018 | 0.047 | 0.014 |
| Mean of age | -0.357 | 0.102 | 0.220 | 0.357 | 0.064 |
| % of female | 0.002 | 0.002 | 0.046 | 0.133 | 0.040 |
| Mean of median household income | -0.259 | 0.019 | 0.130 | 0.259 | 0.064 |
| System size | -0.022 | 0.005 | 0.018 | 0.037 | 0.010 |
| % of white | -0.090 | 0.001 | 0.037 | 0.090 | 0.031 |
| % of black | 0.063 | 0.008 | 0.061 | 0.146 | 0.042 |
| % of Hispanic | 0.165 | 0.077 | 0.136 | 0.187 | 0.036 |
| % of other | -0.132 | 0.027 | 0.071 | 0.134 | 0.036 |

Description of the variables listed in the Table 9 is reported in Table 5 in the Appendix A1.